\documentclass[aps,prl,twocolumn,showpacs]{revtex4}
\usepackage{epsfig}
\usepackage{amsbsy,latexsym}
\usepackage{amsmath}
\usepackage{amssymb, mathrsfs}
\usepackage[mathscr]{eucal}
\usepackage[matrix,frame,arrow]{xypic}\input{Qcircuit}
\def\transp#1{{#1}^\intercal}

\def\<{\langle}\def\>{\rangle}
\def\Tr{\operatorname{Tr}}\def\:{\hbox{\bf :}}
\def\vec#1{{\boldsymbol{#1}}}\def\set#1{{\sf #1}}
\def\dag{\dagger}
\def\geq{\geqslant}\def\leq{\leqslant}
\def\>{\rangle}
\def\<{\langle}
\def\map#1{\mathcal #1}\def\supermap#1{\mathscr #1}
\def\sH{\mathcal{H}}\def\sK{\mathcal{K}}\def\sA{\mathcal{A}}\def\sB{\mathcal{B}}

\def\Span{\set{span}}

\def\Bndd#1{\mathscr B(#1)}
\def\qed{$\,\blacksquare$\par}

\newtheorem{lemma}{Lemma}

\newtheorem{corollary}{Corollary}
\newtheorem{theorem}{Theorem}
\def\Proof{\medskip\par\noindent{\bf Proof. }}

\begin{document}

\title{Transforming quantum operations: quantum supermaps} 
\author{Giulio Chiribella}\email{chiribella@fisicavolta.unipv.it} 
\affiliation{{\em QUIT Group}, Dipartimento di Fisica  ``A. Volta'' and INFM, via Bassi 6, 27100 Pavia, Italy}
\homepage{http://www.qubit.it}
\author{Giacomo Mauro D'Ariano}\email{dariano@unipv.it}
\affiliation{{\em QUIT Group}, Dipartimento di Fisica  ``A. Volta'' and INFM, via Bassi 6, 27100 Pavia, Italy}
\homepage{http://www.qubit.it}
\author{Paolo Perinotti}\email{perinotti@fisicavolta.unipv.it} 
\affiliation{{\em QUIT Group}, Dipartimento di Fisica  ``A. Volta'' and INFM, via Bassi 6, 27100 Pavia, Italy}
\homepage{http://www.qubit.it}
\date{\today}

\begin{abstract}
  We introduce the concept of {\em quantum supermap}, describing the
  most general transformation that maps an input quantum operation
  into an output quantum operation.  Since quantum operations include
  as special cases quantum states, effects, and measurements, quantum
  supermaps describe all possible transformations between elementary quantum objects (quantum systems as well as quantum
  devices).  After giving the axiomatic definition of supermap, we
  prove a realization theorem, which shows that any supermap can be
  physically implemented as a simple quantum circuit.  Applications to
  quantum programming, cloning, discrimination, estimation,
  information-disturbance trade-off, and tomography of channels are
  outlined.
\end{abstract}
 \maketitle

\par The input-output description of any quantum device is provided by the {\em quantum operation}
of Kraus  \cite{kraus}, which yields the most general probabilistic evolution of a quantum state.
Precisely, the output state $\rho_{out}$ is given by the quantum operation $\map{E}$ applied to the
input state $\rho_{in}$ as follows
\begin{equation}
\rho_{out}=\frac{\map{E}\left( \rho_{in}\right)}{\mbox{Tr}\left[\map{E}\left( \rho_{in}\right)
  \right]},\qquad p(\map{E}|\rho_{in}):=\mbox{Tr}\left[\map{E}\left( \rho_{in}\right)
  \right],
\end{equation}
where $p(\map{E}|\rho_{in})$ is the probability of $\map{E}$ occurring on state $\rho_{in}$, when $\map{E}$ is one of
a set of alternative transformations, such as in a quantum measurement.  Owing to its physical
meaning, a quantum operation $\map{E}$ must be a linear, trace non-increasing, completely positive
(CP) map (see, e.g.  \cite{Nielsen2000}).  The most general form of such a map is known as Kraus form
\begin{equation}
\map{E}(\rho)=\sum_j E_j\rho E_j^\dag,
\end{equation}
where the operators $E_j$ satisfy the bound $\sum_jE_j^\dag E_j\leq I$ so that $0\leq
p(\map{E}|\rho_{in})\equiv\Tr[\sum_jE_j^\dag E_j\rho_{in}]\leq 1$.   
 Trace-preserving maps, i.e. those achieving the bound, are a particular kind of quantum operations: they occur deterministically and are 
referred to as {\em quantum channels}.

In general it is convenient to consider two different input and output
Hilbert spaces $\sH_{in}$ and $\sH_{out}$, respectively. In this way,
the concept of quantum operation can be used to treat also quantum
states, effects, and measurements, which describe the properties of
elementary quantum objects such as quantum systems and  measuring devices. Indeed, states can be described as quantum
operations with one-dimensional $\sH_{in}$, i.e. with Kraus operators
$E_j$ given by ket-vectors $ \sqrt{p_j} |\psi_j\> \in \sH_{out}$, thus
yielding the output state $\rho_{out} = \map E(1) = \sum_j p_j
|\psi_j\>\<\psi_j|$.  A {\em quantum effect} $0\le P \le I$
\cite{ludwig} corresponds instead to a quantum operation with
one-dimensional $\sH_{out}$, i.e. with Kraus operators given by
bra-vectors $E_i= \<v_i|$, yielding the probability $p(\map
E|\rho_{in}) \equiv \map{E}(\rho_{in})=\sum_i\<v_i|\rho_{in}|v_i\>=
\Tr[P \rho_{in} ]$ with $P= \sum_i |v_i\>\<v_i|$.  More generally, any
quantum measurement can be viewed as a particular quantum operation,
namely as a quantum-to-classical channel \cite{qc}.

Channels, states, effects, and measurements are all special cases of quantum operations. What about
then considering maps between quantum operations themselves? They would describe the most general
kind of transformations between elementary quantum objects. For example a programmable
channel \cite{NielsenProg} would be a map of this type, with a quantum state at the input and a
channel at the output. Or else, a device that optimally clones a set of unknown unitary gates would
be a map from channels to channels. We will call such a general class of quantum maps \emph{quantum
  supermaps}, as they transform CP maps (sometimes referred to as superoperators) into CP maps.

In this paper we develop the basic tools to deal with quantum
supermaps.  The concept of quantum supermap is first introduced
axiomatically, by fixing the minimal requirements that a map between
quantum operations must fulfill.  We then prove a realization theorem that provides any
supermap with a physical implementation in terms of a simple quantum
circuit with two open ports in which the input operation $\map E$ can
be plugged. This result allows one to simplify the description of
complex quantum circuits and to prove general theorems in quantum
information theory. Moreover, the generality of the concept of
supermap makes it fit for application in many different contexts,
among which quantum programming, calibration, cloning, and estimation of
devices.

To start with, we define the \emph{deterministic supermaps} as those sending channels to channels.
Conversely, a \emph{probabilistic} supermap will send channels to arbitrary trace-non-increasing
quantum operations.  The minimal requirements that a deterministic supermap $\tilde{\supermap{S}}$
must satisfy in order to be physical are the following: it must be {\em i)} linear and {\em ii)}
completely positive. Linearity is required to be consistent with the probabilistic interpretation.
Indeed, if the input is a random choice of quantum operations $\map{E}=\sum_i p_i \map{E}_i$, the
output must be given by the same random choice of the transformed operations
$\tilde{\supermap{S}}(\map{E})=\sum_i p_i\tilde{\supermap{S}}(\map{E}_i)$, and, if the input is the
quantum operation $\map E$ with probability $p$, the output must be the $\tilde{\supermap S} (\map
E)$ with probability $p$, implying $\tilde{\supermap S} ( p \map E) = p \tilde{\supermap S} (\map
E)$.  Clearly, these two conditions imply that $\tilde{\supermap S}$ is a linear map on the linear
space generated by quantum operations.  Complete positivity is needed to ensure that the output of
$\tilde{\supermap{S}}$ is a legitimate quantum operation even when $\tilde{\supermap{S}}$ is applied
locally to a bipartite joint quantum operation, i.e. a quantum operation $\map E$ with bipartite
input space $\sH_{in} = \sH_{in,A} \otimes \sH_{in,B}$ and bipartite output space $\sH_{out}=
\sH_{out,A} \otimes \sH_{out,B}$.  If $\tilde{\supermap S}$ is a supermap transforming quantum
operations with input (output) space $\sH_{in,A}$ ($\sH_{out,A}$), complete positivity corresponds
to require that $\tilde{\supermap S} \otimes \supermap I_{B} (\map E)$ is a CP map for any bipartite
quantum operation $\map E$, $\supermap I_B$ denoting the identity supermap on the spaces labeled by $B$.

In order to deal with complete positivity it is convenient to use the Choi representation
\cite{choi} of a CP map $\map{E}$ in terms of the positive operator $E$ on
$\sH_{out}\otimes\sH_{in}$
\begin{equation}\label{choieq}
E:=\map{E}\otimes\map{I}(|I\>\<I |),
\end{equation}
where $|I\>$ is the maximally entangled vector
$|I\>=\sum_n|n\>|n\>\in\sH_{in}^{\otimes 2}$, $\{|n\>\}$  an orthonormal basis, and $\map{I}$ is the identity
operation. The correspondence $E\leftrightarrow\map{E}$ is one-to-one, the inverse
relation of Eq. (\ref{choieq}) being
\begin{equation}\label{Choi}
\map{E}(\rho):=\Tr_{\sH_{in}}[(I\otimes\transp{\rho}) E],
\end{equation}
where $\transp{}$ denotes transposition in the basis $\{|n\>\}$. In
terms of the Choi operator, the probability of occurrence of $\map E$
is given by $p(\map E | \rho_{in})= \Tr [\transp{\rho_{in}} P]$, where
$P$ is the effect $P := \Tr_{\sH_{out}}[ E]$.  To have unit
probability on any state, a quantum channel must have $P=I$, i.e. its
Choi operator must satisfy the normalization
\begin{equation}\label{NormChoiChannel}
\Tr_{\sH_{out}} [E] = I_{\sH_{in}}~.
\end{equation}

A supermap $\tilde{\supermap{S}}$  maps quantum operations into quantum operations as
$\map{E}'=\tilde{\supermap{S}}(\map{E})$. In the Choi representation, the supermap $\tilde{\supermap{S}}$  induces a linear map $\supermap{S}$ on Choi operators, as
$E'=\supermap{S}(E)$.  Using Eq. (\ref{Choi}), we can get back $\tilde{\supermap{S}}$ from
$\supermap{S}$ as follows 
\begin{equation}\label{uffa}
\map{E}'(\rho)=\tilde{\supermap{S}}(\map{E})(\rho)
=\Tr_{\sK_{out}}[(I\otimes\transp{\rho})\supermap{S}(E)].
\end{equation} 
Of course complete positivity of $\map{E}'$ implies that the map $\supermap{S}$
is positive. On the other hand, it is easily seen that the bipartite structure of a joint
operation $\map{E}$ over a composite system induces a bipartite structure
of the Choi operator $E$. The local application of the supermap $\tilde{\supermap{S}}$---given by $\tilde{\supermap S} \otimes \supermap I (\map E)$---then corresponds to the local application of $\supermap{S}$---given by $\supermap S \otimes \supermap I (E)$---whence $\tilde{\supermap{S}}$ is CP  if and 
only if $\supermap{S}$ is CP. 

Since the correspondence $\tilde{\supermap{S}}\leftrightarrow\supermap{S}$ is one-to-one, in the
following we will focus our attention on $\supermap{S}$. The supermap $\supermap{S}$ sends
positive operators $E$ on $\sH_{out}\otimes\sH_{in}$ to positive operators $\supermap S(E)$ on generally
different Hilbert spaces $\sK_{out}\otimes\sK_{in}$. Complete positivity of $\supermap{S}$ is equivalent to
the existence of a Kraus form 
\begin{equation}\label{SS}
  \supermap{S}(E)=\sum_iS_i E S_i^\dag,
\end{equation}
where $\{S_i\}$ are operators from $\sH_{out}\otimes\sH_{in}$ to $\sK_{out}\otimes\sK_{in}$. 

The following Lemmas provide the characterization of
  deterministic supermaps:
  \begin{lemma} Any linear operator $C$ on $\sH_{out}\otimes\sH_{in}$ such that $\Tr[C E]=1$ for all
    Choi operators $E$ of channels has the form $C=I\otimes\rho$, with $\rho$ on $\sH_{in}$
    satifying $\Tr[\rho]=1$. For $C\geq 0$ one has $\rho\geq 0$.
\end{lemma}
\Proof Consider a Choi operator $E$ with effect $\Tr_{\sH_{out}}[E]=P \le I$.  Upon defining $D :=
\sigma \otimes (I-P)$ for some state $\sigma$ on $\sH_{out}$ we have that $E+D$ is the Choi
operator of a channel, normalized as in Eq.  \eqref{NormChoiChannel}. Since by 
hypothesis $\Tr[C(E+D)]=1$, we have that 
\begin{align}
\Tr[CE]=&1-\Tr[CD]=1-\Tr[C(\sigma\otimes I)]+\Tr[C(\sigma\otimes P)]\nonumber
\\=&\Tr[C(\sigma\otimes P)],
\end{align}
since $\sigma\otimes I$ is the Choi operator of a channel. Therefore,
\begin{equation}
\begin{split}
\Tr[CE]=&\Tr[C(\sigma\otimes P)]=\Tr[P\Tr_{\sH_{out}}[C(\sigma\otimes I)]]\\=&
\Tr[\Tr_{\sH_{out}}[E]\Tr_{\sH_{out}}[C(\sigma\otimes I)]]\\=&
\Tr[\rho P],
\end{split}
\end{equation}
where $\rho:=\Tr_{\sH_{out}}[C(\sigma\otimes I)]$.  Since $\rho$ does not
depend on $E$, the last equality can be rewritten as
$\Tr[CE]=\Tr[(I\otimes \rho) E]$ for all positive $E$, whence
$C=I\otimes\rho$, and in order to have $\Tr[CE]=1$ for all $E$, we must have $\Tr[\rho]=1$. 
Clearly $C\geq 0$ implies $\rho\geq0$.\qed

\begin{lemma}\label{lemma:S*} The supermap $\supermap{S}$ is deterministic iff there
  exists a channel $\map{N}_*$ from states on $\sK_{in}$ to states on $\sH_{in}$ such that, for any state $\rho$ on $\sK_{in}$,
  one has
\begin{equation}\label{S*}
\supermap S_* (I_{\sK_{out}} \otimes \rho) = I_{\sH_{out}} \otimes \map N_*(\rho)~,
\end{equation} 
where $\supermap S_*$ is the dual map of $\supermap S$ defined in
terms of the Kraus form in Eq.~\eqref{SS} by
\begin{equation}
\supermap S_* (O) :=
\sum_i S^\dag_i O S_i~.
\end{equation} 
\end{lemma} 
\Proof One has $\Tr[C\supermap S(E)]=\Tr[\supermap S_*(C)E]$. Consider a positive operator
$C=I\otimes\rho$ on $\sK_{out}\otimes\sK_{in}$, where $\rho$ is a state on $\sK_{in}$. We have that
\begin{equation}
  1=\Tr[C\supermap S(E)]=\Tr[\supermap S_*(C)E],
\end{equation}
for all Choi operators $E$ of channels. According to Lemma \ref{SS}, this implies $\supermap
S_*(I\otimes\rho)=I\otimes\sigma$, where $\sigma$ is a state for any state $\rho$. Since the maps
$\rho\mapsto I\otimes\rho$, $\supermap S_*$ and $I\otimes\sigma\mapsto\sigma$ are all CP, we have
$\sigma=\map N_*(\rho)$, where $\map N_*$ is a CP trace preserving map from states on $\sK_{in}$ to
states on $\sH_{in}$.\qed

Remarkably, the same mathematical structure of Lemma \ref{lemma:S*}
characterizes \emph{semi-causal quantum operations} \cite{ESW}, i.e.
operations on bipartite systems that allow signaling from system $A$
to system $B$ but not viceversa. In our case, this structure
originates from the causality of input-output relations. An equivalent
condition for a supermap to be deterministic is given by the
following:

\begin{lemma}
  The supermap $\supermap{S}$ is deterministic iff there exists an
  identity preserving completely positive map $\map{N}$ such that, for
  any operator $E$ on $\sH_{out}\otimes\sH_{in}$, one has
  \begin{equation}\label{N}
    \Tr_{\sK_{out}}[\supermap{S}(E)]=\map{N}(\Tr_{\sH_{out}}[E]).
  \end{equation} 
\end{lemma}
\Proof This lemma follows from the previous one by considering that
\begin{equation}
\begin{split}
  &\Tr[\rho\Tr_{\sK_{out}}[\supermap
  S(E)]]=\Tr[(I\otimes\rho)\supermap S(E)]\\
  &=\Tr[\supermap S_*(I\otimes\rho)E]=\Tr[(I\otimes \map N_*(\rho))E]\\
  &=\Tr[(I\otimes\rho)(\map I\otimes\map N)(E)]=\Tr[\rho\map
  N(\Tr_{\sH_{out}}[E])],
\end{split}
\end{equation}
for all states $\rho$ on $\sK_{in}$. The map $\map N$ is identity
preserving because it represents $\map N_*$ in the Schr\"odinger
picture.\qed

Eq. \eqref{N} shows that the effect $P'=\Tr_{\sK_{out}}[\supermap
S(E)]$ depends only on the effect $P=\Tr_{\sH_{out}}[E]$, e.~g.
\emph{not} on $\Tr_{\sH_{in}}[E]$.  Basically, this reflects the fact
that, in the input/output bipartition of the Choi operator, the output
must not influence the transformation of the input effect.

\begin{figure}[h]
\newlength{\polowidth}
\newcommand{\poloFantasmaCn}[1]{{{}^{#1}_{\phantom{#1}}}}
\centerline{\large
$$\Qcircuit @C=1.5em @R=.7em{
& \poloFantasmaCn{\sK_{in}} & \multigate{1}{\map{V}} & \qw \poloFantasmaCn{\sH_{in}} &
\gate{\map{E}} & \qw \poloFantasmaCn{\sH_{out}} & \multigate{1}{\map{W}} & \qw
\qw &\qw & \qw \poloFantasmaCn{\sK_{out}}  & \\ 
& & *+<1em,.9em>{\hphantom{\map{V}}} & \qw & \qw \poloFantasmaCn{\sB_{} } & \qw &
\ghost{\map{W}} & \poloFantasmaCn{\sA_{} } \qw & \meter &
\\ {} \gategroup{1}{3}{2}{9}{1.1em}{--} 
}$$}
\caption{Realization scheme for a supermap
  $\tilde{\supermap{S}}$ sending the quantum operation $\map{E}$
  to the quantum operation $\map{E}'=\tilde{\supermap{S}}(\map{E})$,
  here represented by the dashed-boxed circuit.  The input quantum
  operation $\map{E}$ sends states in $\sH_{in}$ to states in
  $\sH_{out}$.  The output quantum operation $\map{E}'$ sends states
  on $\sK_{in}$ to states on $\sK_{out}$.  The supermap is realized by
  two maps $\map{V}=V\cdot V^\dag$ and $\map{W}=W\cdot W^\dag$ made
  by isometries $V$ and $W$ located at the input and at
  the output ports of the quantum operation $\map{E}$, respectively. The two
  isometries are possibly connected by an identity channel on the ancillary
  system with Hilbert space $\sB$.  At the output the ancilla with
  Hilbert space $\sA$ is either measured, each outcome post-selecting a {\em
    probabilistic supermap}, or simply discarded, thus
  realizing a {\em deterministic supermap}.}\label{f:scheme}
\end{figure}
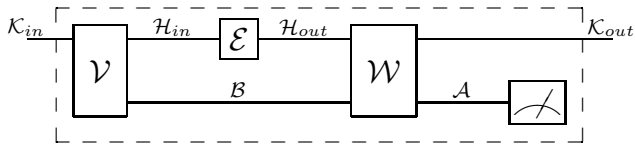

Now we show that deterministic supermaps, so far introduced on a purely axiomatic level, can be
physically realized with simple quantum circuits.  Upon writing a canonical Kraus form for the
completely positive map $\map{N}$ as follows
\begin{equation}\label{NN}
\map{N}(P)=\sum_lN_l^\dag P N_l,
\end{equation}
and substituting the Kraus forms (\ref{SS}) and (\ref{NN}) into Eq. (\ref{N}), one obtains
\begin{equation}\label{topbot}
\begin{split}
&\sum_n(\< k_n|\otimes I)S_i E S_i^\dag (I\otimes |k_n\>)\\
=&\sum_m(\<h_ m|\otimes N_j^\dag) E (N_j\otimes |h_m\>), 
\end{split}
\end{equation}
where $\{|k_n\>\}$ and $\{|k_m\>\}$ are orthonormal basis for $\sK_{out}$ and $\sH_{out}$,
respectively, and identity operators must be considered as acting on the appropriate Hilbert
spaces---$\sK_{in}$ on the top and $\sH_{in}$ on the bottom part of Eq. (\ref{topbot}). Eq.
(\ref{topbot}) gives two equivalent Kraus forms for the same CP map, of which the second
one is canonical (since $\{N_j\}$ is canonical and $\{|h_m\>\}$ are orthogonal). Therefore, there
exists an isometry $W$ connecting the two sets of Kraus operators as follows
\begin{equation}\label{iso}
(\<k_n|\otimes I)S_i=\sum_{mj}W_{ni,mj} (\< h_m|\otimes N_j^\dag),
\end{equation}
with $W^\dag W=I$. Explicitly 
\begin{equation}
W_{ni,mj}:=(\<k_n|\otimes\<a_i|)W(|h_m\>\otimes|b_j\>),
\end{equation}
where $\{|a_i\>\}$ and $\{|b_j\>\}$ are orthonormal basis for two ancillary systems with Hilbert spaces
$\sA$ and $\sB$. From Eq. (\ref{iso}) we then obtain
\begin{equation}\label{So}
S_i = (I\otimes\<a_i|)W(I\otimes Z),
\end{equation}
where 
\begin{equation}\label{Z}
Z=\sum_j|b_j\>\otimes N_j^\dag~.
\end{equation}
Using Eq. (\ref{SS}) we can now evaluate the output Choi operator as follows
\begin{equation}
\supermap{S}(E)=\Tr_\sA[W(I\otimes Z) E (I\otimes Z^\dag)W^\dag].
\end{equation}
Finally, using Eq. (\ref{uffa}) we get
\begin{equation}
\begin{split}
&\map{E}'(\rho)=\Tr_{\sK_{in}} [(I \otimes\transp{\rho})   \supermap S(E) ]\\
&=\Tr_{\sK_{in}\otimes\sA}[(I_{\sK_{out}\otimes\sA}\otimes\transp{\rho})W(I\otimes Z) E (I\otimes Z^\dag)W^\dag]\\
&=\Tr_{\sA}[W(\map{E}\otimes\map{I}_\sB)(V\rho V^\dag)W^\dag],
\end{split}
\end{equation}
where $V=\sum_j|b_j\>\otimes N_j^*$ is the partial transposed of $Z$ (see Eq. (\ref{Z}))
on the second space. Since the map $\map N$ is identity preserving,
$V$ is an isometry, namely $V^{\dag} V = I$.  We have then proved the
following realization theorem
\begin{theorem}\label{maintheorem} Every deterministic supermap
  $\tilde{\supermap{S}}$ can be realized by a four-port quantum circuit where
  the input operation $\map E$ is inserted between two isometries $V$
  and $W$ and a final ancilla is discarded as in Fig. \ref{f:scheme}.
  The output operation $\map E' = \tilde{\supermap S}(\map E)$ is
  given by
\begin{equation}
\tilde{\supermap S}(\map E)(\rho)= \Tr_{\sA}[W(\map{E}\otimes\map{I}_\sB)(V\rho V^\dag)W^\dag].
\end{equation}
\end{theorem}
Since any isometry can be realized as a unitary interaction with an ancilla initialized in some
reset state, the above Theorem entails a realization of supermaps in terms of unitary gates.
However, we preferred stating it in terms of isometries in order to avoid the arbitrariness in the
choice of the initial ancilla state.

We want now to emphasize that deterministic supermaps with $\sH_{in}=
\sK_{in}$ do not preserve in general the probabilities of occurrence
of arbitrary quantum operations: if $\map E$ is not a channel $p(\map
E'|\rho_{in})$ can be different from $p(\map E |\rho_{in})$. This is
clear from Fig.  \ref{f:scheme}, since the input state $\rho_{in}$ is
generally changed by the isometry $V$. Indeed, one can have the
extreme situation in which for every $\rho_{in}$ the isometry $V$
feeds into $\map{E}$ a fixed state on which $\map E$ occur with
certainty, thus transforming the probabilistic quantum operation $\map
E$ into a deterministic one $\map E'$. For the above reason, we will
call {\em probability preserving} those special deterministic
supermaps with $\sH_{in}=\sK_{in}$ that also preserve occurrence
probability for all states, namely which preserve the effect
$P=\Tr_{\sH_{out}}[E]$.  Since the input effect
$P=\Tr_{\sH_{out}}[E]$ is transformed into the output effect
$P'=\Tr_{\sH_{out}}[E']$ by the map $\map N$, we may denote $\map{N}$ as the {\em effect-map} associated to the supermap
$\supermap{S}$, as in Eq. (\ref{N}).  It is immediate to see that the supermap $\map{S}$ is
probability preserving if and only if its effect-map $\map{N}$ is the identity
map.

Up to now we have considered only deterministic supermaps. What about
the probabilistic ones?  By definition a probabilistic supermap
$\supermap{S}$ turns quantum channels into arbitrary
trace-nonincreasing quantum operations. In this case, it is not always
possible to associate an effect-map $\map{N}$ to $\supermap{S}$.
However, a probabilistic supermap $\supermap{S}_1$ is always completed
to a deterministic one by some other supermaps $\supermap{S}_2$,
$\supermap{S}_3$, $\ldots$, which can occur in place of
$\supermap{S}_1$, so that
$\supermap{S}_1+\supermap{S}_2+\ldots=:\supermap{S}$ is deterministic.
Each supermap $\supermap{S}_j$ is completely positive, hence it has a
Kraus form with operators $\{S_k^{(j)}\}$, and all Kraus forms
together provide a Kraus form for the deterministic supermap
$\supermap{S}$. Therefore, since according to Eq. (\ref{So}) each
Kraus term $S^{(j)}_k$ is associated to an outcome of a von Neumann
measurement $\{ P^{(j)}_k= |a^{(j)}_k\>\<a^{(j)}_k | \}$ over the
ancilla with Hilbert space $\sA$, any probabilistic supermap can be
realized by a quantum circuit as in Fig. \ref{f:scheme}, via
postselection induced by a projective measurement $\{ P^{(j)}= \sum_k
P^{(j)}_k \}$ over the ancilla.
\begin{theorem}\label{maintheorem2} Every probabilistic supermap
$\tilde{\supermap{S}}$ can be realized by a four port scheme with measurement as in
Fig.~\ref{f:scheme}, namely
\begin{equation}\label{probS}
\tilde{\supermap{S}}(\map{E})(\rho)=
\Tr_{\sA}[(P \otimes I) W(\map{E}\otimes\map{I}_\sB)(V\rho V^\dag)W^\dag].
\end{equation}
with $V$ and $W$ isometries, and $P$  orthogonal projector over a subspace of the ancillary space $\sA$.
\end{theorem}
We want to stress the generality of the realization Theorem
\ref{maintheorem2}, which can be seen as the analog for probabilistic supermaps (here presented in finite dimensions) of Ozawa's realization theorem for quantum instruments \cite{ozawa}. Indeed Eq.  (\ref{probS}) describes any circuit in
which an input device can be plugged, e.~g. circuits with measurements
performed in different stages, including the possibility of
conditioning transformations on measurement outcomes. Therefore,
whatever the input and the output of the quantum circuit might be
(states, channels, or measurements), the following \emph{delayed
  reading principle} will hold at a fundamental axiomatic level:
\begin{corollary}[Delayed reading principle] Every probabilistic
  quantum circuit is equivalent to a unitary circuit with a single
  orthogonal measurement at the output.
\end{corollary}
Quantum supermaps can be applied to a tensor product of quantum
operations, namely to a set of quantum operations that are not
causally connected (the output of one map is not used as the input for
another map).  Assorted input sets of states, channels, and
measurements can be considered as well, as long as they are not
causally connected.  Differently, if one wants to map an input set of
two causally connected quantum operations, or possibly a memory
channel  \cite{WK}, one needs to move to higher level of supermaps,
namely {\em supermaps of supermaps} \cite{Nota}.  Since the supermap is
CP, one can introduce its Choi operator, and then consider the
physically admissible mappings.  In this way, one can build up a whole
hierarchy of supermaps by considering the completely positive maps
acting on the Choi operators of the lower level. An efficient
diagrammatic approach to treat this problem is provided in Ref.
 \cite{QCA} by introducing the notion of \emph{quantum comb}.  The
normalization condition for such higher-level supermaps has a
recursive form, entailing the causal structure of input-output
relations.

\par Before concluding, we outline list here some remarkable \emph{ante litteram} examples of supermaps as well as some novel applications of this theoretical tool: 

\par {\em 1. Quantum Compression of Information and Error Correction.}
The realization scheme of Fig. \ref{f:scheme} entails as a special
case the coding/decoding scheme at the basis of quantum error
correction and information compression.  Schumacher's information
compression \cite{Schumacher} is a beautiful ante litteram example of
supermap, which turns a noiseless communication channel on a smaller
system into a channel that reliably transfers states in a larger
Hilbert space. Similarly, also error correction can be seen as a
supermap, now turning a noisy channel on a larger Hilbert space into a
noiseless channel acting on a smaller space. In both cases the supermap
is given by the insertion of the input channel $\map{E}$ between two
deterministic channels $\map{C}$ and $\map{D}$ (the {\em coding} and
{\em decoding} maps, respectively), namely $\tilde {\supermap S} (\map
E) = \map D \map E \map C$,  with the additional constraint that the
ancilla $\sB$ in Fig. \ref{f:scheme} must be one-dimensional.

\par {\em 2. Cloning of transformations.}
An interesting application of quantum supermaps is the optimal cloning
of transformations, instead of states. For example, an optimal $1\to
2$ cloner of unitary transformations would be a four-port circuit that
turns an unknown unitary channel $\map{U}=U\cdot U^\dag$ into a
channel $\tilde{\supermap S} (\map U)$ that maximizes the average channel fidelity with
the bipartite channel $\map{U} \otimes \map U$.  This device has been
recently studied in Ref.  \cite{ourclon}, and has optimal global
fidelity ${F}=(d+\sqrt{d^2-1})/d^3$, surpassing the value achievable
by any  classical cloning scheme. The non-classical
performances of the cloning circuit essentially depend on the
possibility of entangling system $\sH_{in}$ with the ancilla $\sB$ in
Fig. \ref{f:scheme}  \cite{ourclon}.  It is rather intriguing to investigate the possible cryptographic connections of the problem of optimally cloning unitary channels, which appear to be an harder task than cloning pure quantum states.

\par {\em 3. Discrimination/estimation of channels and memory channels.} A
probabilistic supermap $\tilde{\supermap S}$ with one-dimensional $\sK_{in}$ and
$\sK_{out}$ sends a quantum operation $\map E$ into a probability $p= \tilde{\supermap S}(\map E)$.
In this case the Kraus operators $S_i$ are given by bra-vectors $\<v_i|$ with $|v_i\> \in \sH_{out} \otimes \sH_{in}$, and Eq.
(\ref{SS}) yields $p= \sum_i \<v_i| E |v_i\> = \Tr[E P]$ where $P:= \sum_i |v_i \>\<v_i|$.  A set of such
probabilistic supermaps $\{\tilde{\supermap S^{(j)}}\}$ that sums up to a
deterministic supermap $\tilde{\supermap S} = \sum_j \tilde{\supermap S}^{(j)}$ plays for channels of the same role that a POVM plays for states: for any channel
$\map E$, the supermap $\tilde{\supermap S}^{(j)}$ gives a probability 
\begin{equation}
p_j = \tilde{
\supermap S}^{(j)}(\map E) = \Tr[E P_j]
\end{equation} 
with $p_j \ge 0$, and $\sum_j p_j =1$.  The normalization of probabilities is ensured by the normalization condition of Eq. (\ref{S*}), which here reads
\begin{equation}
\sum_{j}  P_j= I_{\sH_{out}} \otimes \sigma~,
\end{equation} 
$\sigma = \map N_* (1)$ being a quantum state on $\sH_{in}$. This set
of probabilistic supermaps, completely specified by the operators
$\{P_j \ge 0\}$, describes the most general setup one can devise in
order to test a given property of a quantum channel, and can be used
to discriminate between two or more channels, or else to estimate a
signal encoded into a parametric family of channels and quantum
operations. Such a set of probabilistic supermaps is a particular case
of \emph{quantum circuit tester} introduced in Refs. \cite{QCA,
  unitdiscr} to treat the discrimination of causally ordered sequences
of channels and the discrimination of memory channels. We notice that
the particular case of probabilistic supermaps treated in this paragraph has been
independently introduced in Ref. \cite{mario} under the name
\emph{process POVM (PPOVM)}.
\par {\em 4. Information-disturbance trade-off for quantum operations.}
When the spaces $\sK_{in}$ and $\sK_{out}$ are non-trivial, a set of
probabilistic supermaps $\{ \tilde{\supermap S}^{(j)}\}$ summing up to a
deterministic one provides for channels the analog of an
instrument. Differently from a tester, which has only
classical output (the outcome $j$), the output of such a
supermap is both a classical outcome and an output quantum operation.
In this setting, supermaps provide the opportunity to address the completely new
problem of  information-disturbance trade-off for quantum channels.
For example, we may try to estimate a completely unknown unitary $U$,
producing at the same time a channel that is the most possibly similar
to $U$. Similarly to the problem of cloning quantum channels, the
information-disturbance trade-off rises the intriguing possibility of
new cryptographic protocols based on channels instead of states. 

\par {\em 5. Quantum Tomography of devices.} 
An interesting example of supermap is also that corresponding to
tomography of quantum devices based their local application on bipartite states  \cite{faith, debby, calib}. Tomography of quantum operations is based on the
supermap that sends an input operation $\map E$ into an output state
$\tilde{\supermap S}(\map E)=\map{E}\otimes\map{I}(F)$ where $F$ is a faithful state on
$\sH_{in}^{\otimes 2} $ \cite{faith}, so that the output state is in
one-to-one correspondence with the input operation.  Note that, in order
to have such a one-to-one correspondence, the map 
$\supermap S (E) = \sum_i S_i E S_i^\dag$ must be invertible, namely
$\supermap S (E) =0$ if and only if $E=0$.  Tracing Eq. (\ref{SS}) with an arbitrary operator $O$ on  $\sK_{out} \otimes \sK_{in}$,  one can easily see that invertibility of $\supermap S$  is equivalent to the
condition 
\begin{equation}\label{Invert}
\Span \{ \supermap S_* (O) ~|~ O \in \Bndd {\sK_{out} \otimes \sK_{in}} \} = \Bndd {\sH_{out} \otimes \sH_{in}}~,
\end{equation}
where $\Span$ denotes the linear span of a set, and $\Bndd {\sH}$ the
set of all operators on $\sH$. Since $\tilde{\supermap S}$ sets an invertible
correspondence between operations and states, one can perform an
informationally complete measurement on the output state to completely
characterize it.  Note that, using probabilistic
supermaps we can also combine the deterministic map $ E \mapsto
\supermap S(E)$ and the infocomplete measurement in a single object, introducing the notion of \emph{informationally complete
  tester} (see also Ref. \cite{mario}), which is a tester with the property that the mapping $E
\mapsto \{ p(j|E) = \Tr[P_j E]\}$ is invertible. In this case, the
invertibility condition of Eq. (\ref{Invert}) becomes
\begin{equation}
\Span\{ P_j\} = \Bndd {\sH_{out} \otimes \sH_{in}}~.
\end{equation}
 As regards tomography of a POVM $\vec
P=\{P_n\}$, this can be identified with the quantum-to-classical
channel $\map{E}_{\vec P}(\rho) =\sum_n\Tr[P_n\rho]|n\>\<n|$, and the
above scheme applies as well.
\par {\em 6. Programmable devices.} Programmable quantum channels \cite{NielsenProg} and
measurements \cite{dp} are a remarkable example of supermaps. 
Impossibility of ideal programmability suggests to study optimal
programmability. One could consider either deterministic-approximate
case \cite{dp,dpc,zimapp}, or the probabilistic-exact one
\cite{dubu,berghill,zibu}. In these applications an input state
$\sigma$ (the program) is turned into a channel $\map E_{\sigma}$ or
into a measurement (POVM) $\vec P_{\sigma} = \{P_{\sigma, j}\}$.  For
channels the supermap is given by $\map E_{\sigma}
(\rho)=\tilde{\supermap{S}}(\sigma)
(\rho)=\Tr_2[U(\rho\otimes\sigma)U^\dag]$, where $U$ is a unitary
interaction.  For programmable POVMs one has the set of probabilistic
supermaps $\{\tilde {\supermap S}^{(j)}\}$ such that $P_{\sigma,j}=
\tilde {\supermap S}^{(j)} (\sigma)= \Tr_2[E_j (I\otimes \sigma)]$,
where $\{E_j\}$ is a joint POVM. Equivalently, regarding the POVM as a
channel from states to classical outcomes, one has the set of
probabilistic supermaps $\{\tilde {\supermap S}^{(j)}\}$ with
one-dimensional $\sH_{in}$ and $\sK_{out}$ so that $p(j|\rho)= \tilde
{\supermap S}^{(j)} (\sigma) (\rho) = \Tr[P_{\sigma,j} \rho]= \Tr[E_j
(\rho \otimes \sigma)]$, where $\{E_j\}$ is a joint POVM.

\par In conclusion, we have introduced the concept of {\em quantum supermap}, as a tool to describe
all possible transformations between elementary quantum objects, i.e. states, channels, and
measurements, with numerous applications to quantum information processing, cloning, discrimination,
estimation, and information-disturbance trade-off for channels, tomography and calibration of
devices, and quantum programming. A realization theorem has been presented, which shows that any
abstract supermap can be physically implemented as a simple quantum circuit. The generality of the
concept of supermap, describing any quantum evolution, allows one to use it as a tool to formulate
and prove general theorems in quantum information theory and quantum mechanics, and to efficiently
address an large number of novel applications.

\par {\em Acknowledgments.---} We thank D. M. Schlingemann and D.
Kretschmann for many useful discussions about semi-causality and
memory channels during the development of this work. We thank an
anonymous referee for valuable suggestions in the revision of the
original manuscript. This work has been founded by Ministero Italiano
dell'Universit\`a e della Ricerca (MIUR) through PRIN 2005.


\begin{thebibliography}{99}
\bibitem{kraus}  K. Kraus, {\em States, Effects, and Operations} (Springer-Verlag,
Berlin,1983).
\bibitem{Nielsen2000} I. L. Chuang and M. A. Nielsen, {\em Quantum
    Information and Quantum Computation} (Cambridge University Press,
  Cambridge, 2000).
\bibitem{ludwig} G. Ludwig, {\em An Axiomatic Basis for Quantum Mechanics I: Derivation of Hilbert
  Space Structure} (Springer, Berlin 1985).
\bibitem{qc} A. S. Holevo, Russ. Math. Surveys {\bf 53}, 1295 (1998);  C. King, J. Math. Phys. {\bf 43}, 1247 (2002).
\bibitem{NielsenProg}  M. A. Nielsen and I. L. Chuang, Phys. Rev. Lett. {\bf 79}, 321 (1997).
\bibitem{choi} M.-D. Choi, Lin. Alg. Appl. {\bf 10} 285 (1975).
\bibitem{ESW} T. Eggeling, D. Schlingemann, and R. F. Werner,  Europhys. Lett. {\bf 57}, 782 (2002). 
\bibitem{WK}  D. Kreschmann and R. F. Werner, Phys. Rev. A. {\bf 72} 062323 (2005). 
\bibitem{QCA} G. Chiribella, G. M. D'Ariano, and P. Perinotti,  arXiv:0712.1325. 
\bibitem{ozawa} M Ozawa, J. Math. Phys. {\bf 5}, 848 (1984).  
\bibitem{Nota} Two causally connected quantum operations are indeed a four-port circuit in which an input device can be plugged, i.e. referring to Fig. \ref{f:scheme} they are a supermap with one-dimensional ancilla space $\sB$.
\bibitem{Schumacher} B. Schumacher, Phys. Rev. A {\bf 51}, 2738  (1995).
\bibitem{ourclon} G. Chiribella, G. M. D'Ariano, and P. Perinotti, arXiv:0804.0129.
\bibitem{unitdiscr} G. Chiribella, G. M. D'Ariano, and P. Perinotti,  arXiv:0803.3237.
\bibitem{mario} M. Ziman, arxiv:0802.3862.
\bibitem{faith} G. M. D'Ariano and P. Lo Presti, Phys. Rev. Lett. {\bf
    91} 047902 (2003)
\bibitem{debby} D. Leung, J. Math. Phys. {\bf 44}, 528 (2003).
\bibitem{calib} G. M. D'Ariano, L. Maccone, P. Lo Presti, Phys. Rev.
  Lett. {\bf 93}, 250407 (2004).
\bibitem{dp} G. M. D'Ariano and P. Perinotti, Phys. Rev. Lett. {\bf 94}, 090401 (2005).
\bibitem{dpc} G. M. D'Ariano and P. Perinotti, in Quantum Probability
  and Infinite Dimensional Analysis {\bf 20}, ed. by L. Accardi, W.
  Freudenberg, and M. Schurmann (World Scientific, Singapore, 2007),
  pag. 173.
\bibitem{zimapp} M. Hillery, M. Ziman, and V. Bu\v zek, Phys. Rev. A
  {\bf 73}, 022345 (2006).
\bibitem{dubu} M. Du\v sek, V. Bu\v zek, Phys. Rev. A {\bf 66}, 022112
  (2002).
\bibitem{berghill} J. A. Bergou and M. Hillery, Phys. Rev. Lett. {\bf
    94}, 160501 (2005).
\bibitem{zibu} M. Ziman and V. Bu\v zek, Phys. Rev. A {\bf 72}, 022343
  (2005). 
\end{thebibliography}
 \end{document}